# MACROSTATE PARAMETER AND INVESTMENT RISK DIAGRAMS FOR 2008 AND 2009

Anca GHEORGHIU and Ion SPÂNULESCU[*]

***Abstract.*** *In this paper are made some considerations of the application of phenomenological thermodynamics in risk analysis for the transaction on financial markets, using the concept of economic entropy and the macrostate parameter introduced by us in a previous works [15,16]. The investment risk diagrams for a number of Romanian listed companies in 2008 and 2009 years were calculated. Also, the evolution of the macrostate parameter during financial and economic crisis in Romania are studied.*

## 1. Introduction

During the first decade of 21$^{st}$ century a new border science between physics and economy, named econophysics, was emerged. At the beginings, econophysics used mostly statistical physics and mathematics methods to characterize and model a range of economic and financial processes, particularly in capital markets, income distribution, interests, and so on (see for example [1-10]).

In the last years, many researchers included in their papers some models based on analogies between the economical phenomena and phenomena from other fields of physics such as thermodynamics, electricity, spectroscopy, phase transitions physics, reliability theory and so on [11-16].

In this paper, for the analysis of the risk in the financial market transaction, is appealing to the phenomenological thermodynamics methods and to the statistical physics results concerning statistical interpretation of the entropy and the equilibrium conditions of the physical systems.

In our previous works [15,16] some considerations upon economical information value were made and used in order to introduce a new econophysics and technical index for the technical analysis of the stock-market diagnostic named macrostate parameter. With the help of this defined macrostate

[*] Hyperion University of Bucharest, 169 Calea Călăraşilor, St., Bucharest, Romania.

47

parameter, the investment risk diagrams for a number of Romanian companies listed at the stock-market in 2006 and 2007 have been established.

In this paper, using the same macrostate parameter and the economic development statistics, the investment risk diagrams for Romanian companies listed in years 2008 and 2009 have been calculed.

## 2. The Macrostate Parameter

As it was shown in the previous papers [15,16], besides its intrinsic value, characterized by the utility value and trade value, an information type value which determines the denomination, role and importance of any product or service, were considered.

In case when the number of information is large it can be spoken about an information beam with dual aspect, similar to a photon beam or other elementary particles (electrons, protons etc.) characterized by a determined motion mass, impulse, energy etc. [16].

As it was shown in [16] taking into account the above consideration the shares as well as money or economic-financial information about the type volume, or prices of the shares are compared to the particles from an information beam rather by similitude and not by identification with objects from the physical reality. In this case as it was shown in the papers [15,16], for a more complete understanding of the stock evolution from the point of view of the price and transacted volumes, the product **price × transacted volume** can be assimilated to the impulse of a particle (which symbolize the respective financial information) defined by the product $pV$ similar to the impulse $p = mv$ of a particle of mass $m$ and speed $v$. Such an index can deliver ampler useful information regarding the "inertia" degree or stability of an asset (shares, financial instruments etc.) than the price, $p$, or the transacted volume, $V$, taken separately.

In the like manner, all the information from the financial market can be assimilated with the particles of a gas of impulses $p = mv$ confined in a precinct ("financial boiler") that is the very capital market (spot markets, forward markets etc.). In this situation it is plausible enough to apply the same principles, laws and results from thermodynamics, kinetic-molecular or statistical physics to describe the assemble of particles states – called microstates – in which the particles that symbolize the information about the shares (or other financial instruments) from the virtual precinct which can exist at various moments. Every "particle-information" contained in the financial boiler (virtual precinct) is characterized – in a first phase – by the product price × volume of transacted shares, i.e. by the parameter $a = pV$ as we have seen above.

After a determined time, as a result of the succession of a numerous microstates which appear because of the agitation and the mixture of the constituent particles, the system reach an equilibrium state which is a **macrostate** that can be described by measurable macrostate parameters [17].

By financial (or economic) macrostate we understand the assembly of all the information and decisions materialized in the share price and the transacted volume (individual and for all the day, hour or minute of transaction) when we refer to a quoted emitter, or to the quote of the stock-market index.

So, if we consider $a = pV$ parameter, a microstate for the capital market is given by the assemble of the price and share's volume situation at a very moment $t$. Starting from the definition of entropy $S = k \ln W$ in thermodynamics, we can introduce a similar parameter named **macrostate parameter**, for the financial markets:

$$P_M = k_B \ln W_B \tag{1}$$

where $W_B$ represents a probability in succeeding a new microstate of the stock-market and $k_B$ is a constant which is specific for that stock-exchange market and for that type of transacted share.

As it was shown in the papers [15,16], the macrostate parameter $P_M$ of the financial markets is given by [15,16]:

$$P_M = \frac{1}{N} \sum_{t=1}^{N} \frac{a_t - a_{t-1}}{a_{t-1}} = S_e \tag{2}$$

where $N$ is the number of microstates and $\frac{a_t - a_{t-1}}{a_{t-1}}$ represents the normalized volatility given by [15,16]:

$$Vol_n = \frac{P_t V_t - p_{t-1} V_{t-1}}{P_{t-1} V_{t-1}} \tag{3}$$

with: $a_t = p_t V_t$ and $a_{t-1} = p_{t-1} V_{t-1}$.

In equation (2), $S_e$ represents the economic entropy, being determined and equal with the macrostate parameter $P_M$ [16].

As a general conclusion, in the papers [15,16] in was shown that from the econophysics point of view, the macrostate parameter $P_M$ is a measure of the disorder (entropy) from the capital market, being similar to the thermodynamic entropy from physics. Its size shows directly the uncertainty degree from the economic (financial) point of view for various transacted companies, being able to constitute a measure of the attractive degree of assets on the medium or long term.

As in paper [16], where we established the investment risk diagrams for Romanian companies listed at the stock market in 2006 and 2007, in this paper, the investment risk diagrams for Romanian companies listed in 2008 and 2009 using the state parameter $P_M$ have been calcuated (see Fig. 1 and Fig. 2).

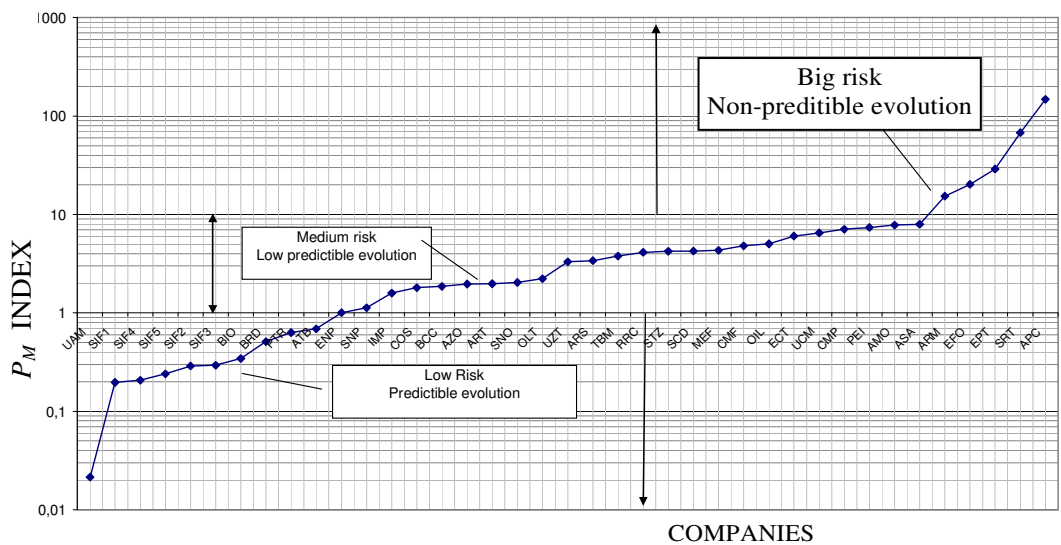

**Figure 1.** Investment risk diagram for 40 romanian companies quoted at Bucharest Stock-Exchange Market (BVB) during 2008 year.

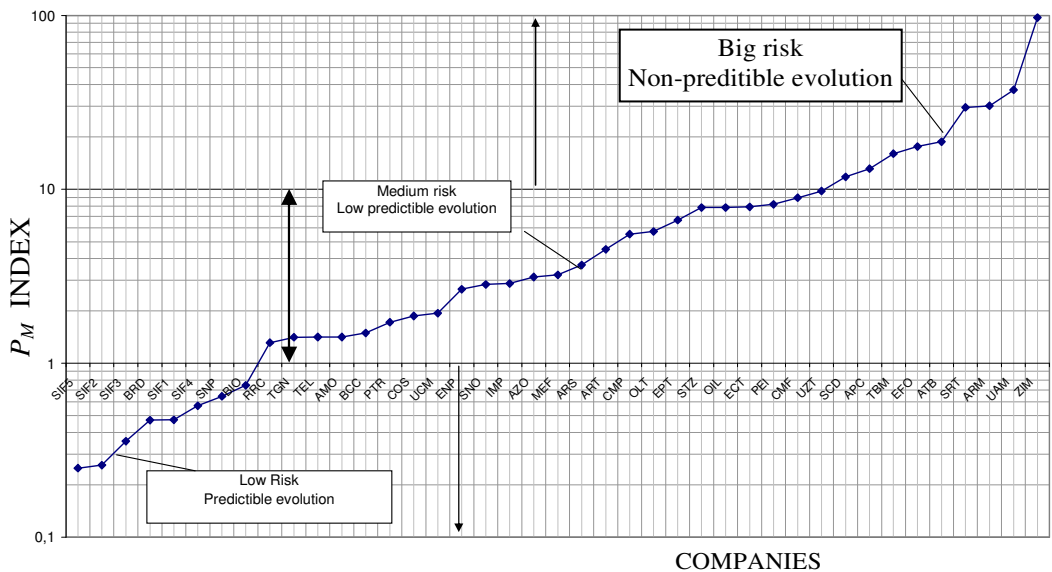

**Figure 2.** Investment risk diagram for 40 romanian companies quoted at Bucharest Stock-Exchange Market (BVB) during 2009 year.

## 3. The Influence of the Economic and Financial Crisis on the Macrostate Parameter

As expected, each listed company suffered the impact of the economic and the financial crisis triggered in 2007-2008. The question is: How will evolve the macrostate parameter $P_M$ for companies during the economic and the financial crisis which in Romania has reached its zenith in years 2008-2009?

For the energy or the petroleum companies or for the producing related equipment companies (OIL, PEI etc.) is found that this parameter greatly increased, reaching a very clearly pronounced maximum that was maintained in 2008-2009 (Fig. 3 and Fig. 4).

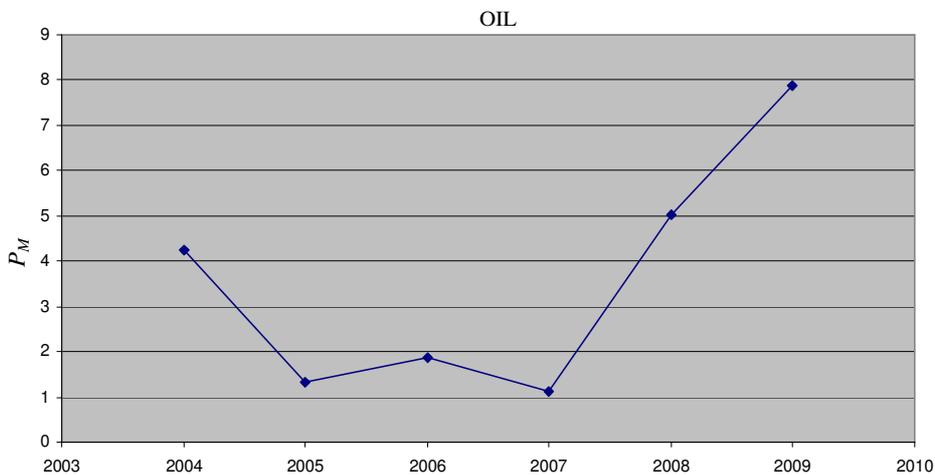

**Figure 3.** Time dependence of $P_M$ for OIL Company during 2004-2009 years.

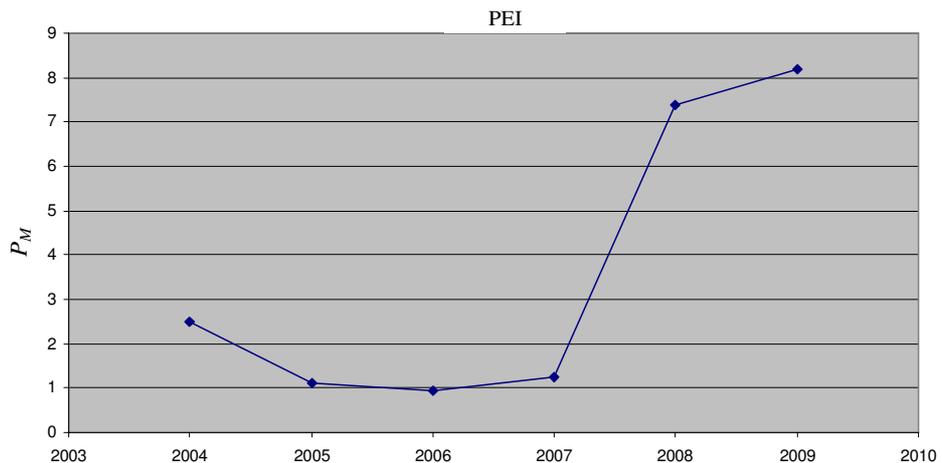

**Figure 4.** Time dependence of $P_M$ for Petrol-Export-Import Company during 2004-2009 years.

Other companies, more stable like BRD, COS, SNP etc., which were subjected to the infusion of capital or to the capital increase etc. proved to have a normal behavior, without high peak values of the parameter $P_M$. The variation of $P_M$ for these companies is given in figures 5, 6 and 7.

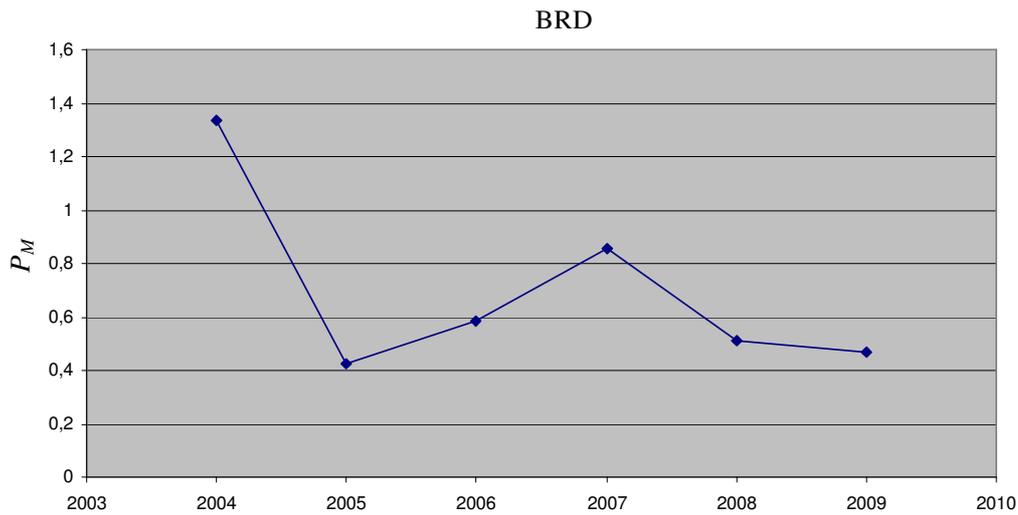

**Figure 5.** Time dependence of $P_M$ for BRD Societé Generale during 2004-2009 years.

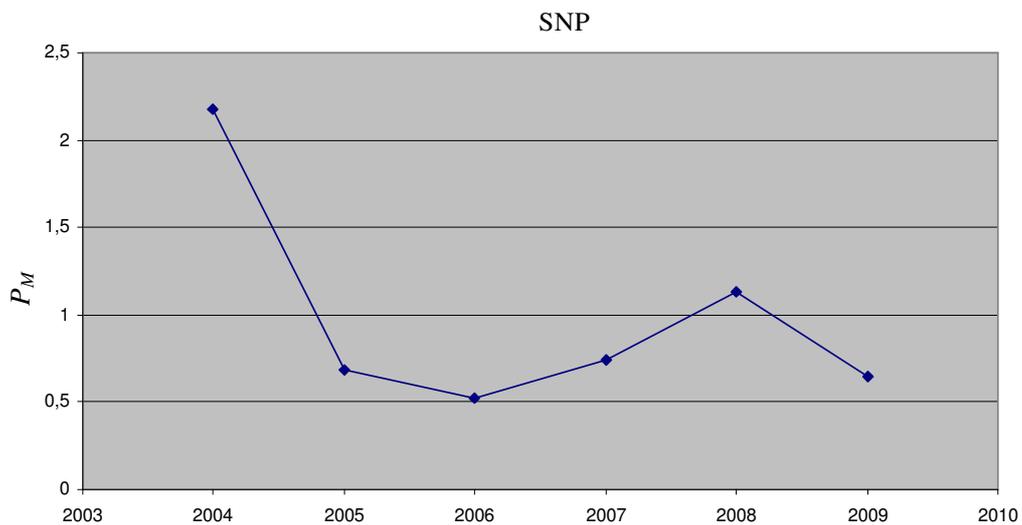

**Figure 6.** Time dependence of $P_M$ for SNP Company during 2004-2009 years.

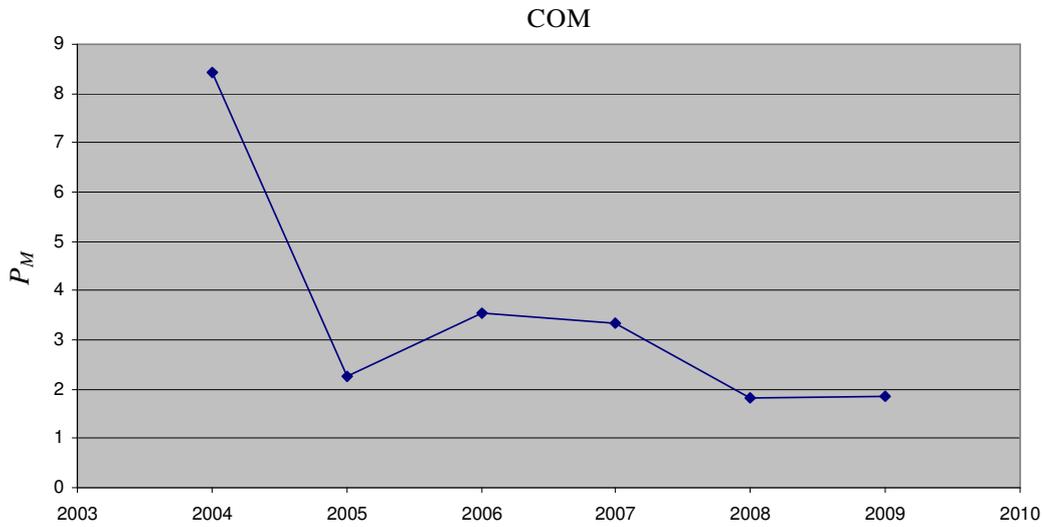

**Figure 7.** Time dependence of $P_M$ for Compa-Sibiu Company during 2004-2009 years.

Thereby, as well as other financial or economic indexes, the macro-state parameter $P_M$ proved extremely sensitive to the economic and the financial condition changes, and properly reflects the condition and the degree of uncertainty for the companies that have presented obvious difficulties during of crisis and/or recession so as shown in figure 3 and figure 4.

## 5. Conclusions

In this paper we continue the investigations over the application of physics methods and especially of phenomenological thermodynamics for business processes modeling, especially in the capital markets field.

Using the economic entropy concept or the macrostate parameter introduced in previous papers [15,16], the investment risk diagrams for a number of Romanian companies listed in years 2008 and 2009 have been calculated. The macrostate parameter evolution during the economic and financial crises for some Romanian companies were also studied.